\documentclass[aps,showpacs,amssymb,floatfix,pra,notitlepage,showkeys]{revtex4-1}

\usepackage{graphicx}
\usepackage{amssymb,amsfonts,amsmath,bm}

\renewcommand{\r}{\mathbf{r}}
\newcommand{\hr}{\widetilde{r}}
\newcommand{\hR}{\widetilde{R}}
\newcommand{\hx}{\widetilde{x}}
\newcommand{\hz}{\widetilde{z}}
\newcommand{\hl}{\widetilde{\ell}}
\newcommand{\eff}{\text{eff}}
\usepackage[utf8x]{inputenc}

\newlength{\GraphicsWidth}
\begin{document}

\title{A two-dimensional one component plasma and a test charge : polarization effects and effective
potential}
\author{Gabriel T\'ellez}
\affiliation{Departamento de F\'{\i}sica, Universidad de Los Andes,
A.A.~4976, Bogot\'a, Colombia}
\author{Emmanuel Trizac}
\affiliation{Laboratoire de Physique Th\'eorique et Mod\`eles Statistiques, 
UMR CNRS 8626, Universit\'e Paris-Sud, 91405 Orsay, France}

\begin{abstract}
We study the effective interactions between a test charge $Q$ and a one-component
plasma, i.e. a complex made up of mobile point particles with charge $q$, and a uniform
oppositely charged background. The background has the form of a flat disk, in which the 
mobile charges can move. The test particle is approached perpendicularly to 
the disk, along its axis of symmetry. All particles interact by a logarithmic
potential. The long and short distance features of the effective potential
--the free energy of the system for a given distance between $Q$ and the disk--
are worked out analytically in detail. They crucially depend on
the sign of $Q/q$, and on the global charge borne by the discotic complex,
that can vanish.
While most results are obtained at the intermediate coupling $\Gamma \equiv \beta q^2 = 2$ ($\beta$
being the inverse temperature), we have also investigated situations with stronger 
couplings: $\Gamma=4$ and 6. We have found that at large distances,
the sign of the effective force reflects subtle details of the charge 
distribution on the disk, whereas at short distances, polarization effects
invariably lead to effective attractions.

\end{abstract}


\maketitle

%%%%%%%%%%%%%%%%%%%%%%%%%%%%%%%%%%%%%%%%%%%%%%%%%%%%%%%%%%%%%%%%%%%%%%%%%%%%%%%
\section{Introduction}
Electric charges are ubiquitous in the colloidal domain, and often a major
player shaping the behaviour of soft matter systems. Counter-intuitive phenomena
often ensue, such as overcharging (charge inversion) or effective attraction between like-charged
macro-ions \cite{GBPP00,HL00,GNS02,L02,L05,M09,NKNP10}. 
To rationalize such observations that are the
fingerprints of correlation effects, simplified models are welcome, that should 
furthermore be treated beyond the mean-field level \cite{T00}. Interestingly, the 
physics of strongly coupled charged systems has witnessed relevant progress in the
last 15 years \cite{RB96,S99,L02,MN02,BKNN05,ST11}, while the study in the weak coupling
limit where mean-field arguments hold, started about 100 years ago
\cite{G10,C13}. The study of intermediate Coulombic couplings, though, 
appears more elusive \cite{BAO04,CW06,S06,BMP10}
and will be the focus of our interest in the present paper.

\begin{figure}[htb]
  \centering
\includegraphics[width=5cm]{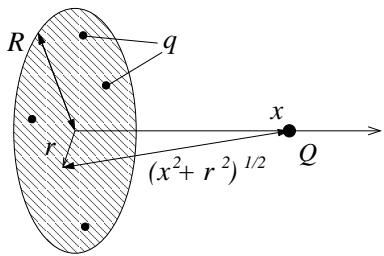}  
\caption{\label{fig:OCPdisk-charge}
Definition of the system under consideration: a disk $\mathcal{D}$ with fixed and uniform 
background charge (hatched area), in which $N$ mobile oppositely
charged particles, shown by the bullets, are free to move ($N=4$ in the figure). 
The total charge of the background is
$-N_b q$, while each mobile ion bears a charge $q$. The total
charge on the disk (background plus free ions) is therefore $(N-N_b)q$.
A test ion with charge $Q$
approaches the disk along the symmetry axis shown, defining
$x$-coordinate ($x=0$ when the test charge lies on the disk,
at the center.)}
\end{figure}

The system under scrutiny here is a variant of Thomson's plum pudding model
(see \cite{T1904,BP94,CCRT11} and references therein), 
also referred to as the One Component Plasma \cite{DL74,J81,L02}.
Point particles with charge $q$ are embedded in a two-dimensional flat disk $\mathcal{D}$ 
of radius $R$. In addition, a uniformly charged background is present in the disk region
(see Fig. \ref{fig:OCPdisk-charge}). 
While the charged background is fixed,
the particles are free to move in $\mathcal{D}$. They interact through a log potential,
the form taken by Coulomb law in two dimensions. The relevant coupling parameter
is $\Gamma = \beta q^2$, where $\beta$ is the inverse temperature. 
At small $\Gamma$ (formally $\Gamma \to 0$), 
the Poisson-Boltzmann mean-field description holds \cite{Hunter}
\footnote{It is straightforward to check that in the globally neutral case $N=N_b$,
the mean-field solution is trivial, with a vanishing electrostatic potential,
and a particle density that compensates for that of the background. This is a
consequence of the confinement in $\mathcal{D}$ imposed to the charges. If the mobile
charges are allowed to leave the uniformly charged disk and explore the whole
2D plane, the mean-field solution becomes non trivial --the constant electrostatic potential
can by no means provide a solution to the problem-- and has been studied 
in \cite{WB57,CMTR09}. We come back to this modified ``unbounded'' model
in our concluding section.}.

As is often the case for 2D Coulombic problems, the coupling parameter $\Gamma=2$
lends itself to an exact analytical treatment, see Refs \cite{J81,AJ81,S04}.
The goal is here to extend the exact analysis at $\Gamma=2$ to investigate the 
interactions between the disk bearing the mobile charges,
%that can be seen as a simple model of a colloid,
and a test charge $Q$ that is approached perpendicularly to the disk,
along the axis of symmetry (see Fig. \ref{fig:OCPdisk-charge}).
We shall assume that $Q$ and all other charges (mobile + background) interact 
through a log potential. Since $Q$ explores a third additional direction 
compared to those in which the point $q$-charges and the background
disk are confined, the choice of such a potential can be questioned: it does not correspond 
to the solution of Poisson's equation in three dimensions. 
This is however the price for obtaining analytical results, that shed light 
on phenomena at work in more realistic systems. In particular,
we will be interested in the effective interactions between the fixed charge $Q$
and other charges, that can be seen as mimicking a colloid (the uniform background),
dressed by a double-layer of counterions (the mobile $q$-charges). 

The model will be defined in section \ref{sec:general}, where the
theoretical tools will also be introduced. In section \ref{sec:long-distance},
the features of the effective potential at large distances will be addressed. 
While most of
the present analysis pertains to the $\Gamma=2$ case, other couplings 
will be addressed (namely $\Gamma=4$ and 6, corresponding to smaller 
temperatures). Then, the emphasis will be in section \ref{sec:short}
on short range correlations that, through polarization effects, 
rule the short distance behaviour of the effective potential. Conclusions will be
drawn in section \ref{sec:concl}. 

%%%%%%%%%%%%%%%%%%%%%%%%%%%%%%%%%%%%%%%%%%%%%%%%%%%%%%%%%%%%%%%%%%%%%%%%%%%%%%
\section{Model and general formalism}
\label{sec:general}

The system is a one-component plasma~\cite{DL74,J81,AJ81,ZW06} on a
disk $\mathcal{D}$ of radius $R$ with $N$ mobile point charges $q$,
and a fixed background charge density $\rho_b=-q n_b$.  The system can
be globally charged, since $N_b = \pi R^2 n_b$ can be different from
$N$. The charged particles interact with the two-dimensional
logarithmic Coulomb potential,
\begin{equation}
  \label{eq:vCoul}
  v_{c}(\r_i,\r_j)=-\ln \frac{|\r_i-\r_j|}{\ell}
  \,,
\end{equation}
for two particles located at $\r_i$ and $\r_j$ on the disk ($\ell$ is
an arbitrary length).
The interaction potential between a charge $q$ located at $\r$ and the
background consequently reads
\begin{eqnarray}
  v_b(r)&=&\int_{\mathcal{D}}
  q \rho_b v_{c}(\r,\r')\,d^2\r'
  \nonumber\\
  &=&
  \frac{\pi n_b q^2}{2}
  \left[
    r^2-R^2 \left(1-2\ln\frac{R}{\ell}\right)
  \right]
  \,,
  \label{eq:vb}
\end{eqnarray}
where $r=|\r|$.
The disk is in thermal equilibrium with a heat bath at an inverse
temperature $\beta=1/k_B T$.
We consider now that a particle with charge $Q$ approaches the disk
from the axis normal to the disk that passes through its center, see
figure~\ref{fig:OCPdisk-charge}. The charge $Q$ is held fixed at a distance $x$
from the disk. The interaction potential between this charged particle
and a charge from the disk located at $\r$, will also be taken 
logarithmic:
\begin{equation}
  \label{eq:vQp}
  v_Q(x,r)=-\ln\frac{\sqrt{x^2+r^2}}{\ell}
  \,.
\end{equation}
It will be convenient to use rescaled lengths $\hr=\sqrt{\pi n_b} r$,
$\hx=\sqrt{\pi n_b} x$, $\hl=\sqrt{\pi n_b} \ell$, etc. With such a 
choice, the rescaled disk radius is $\hR = \sqrt{N_b}$.
The
interaction potential between the background and the approaching
particle is 
\begin{eqnarray}
  \label{eq:VQb}
  V_{Qb}(x)&=&\int_{\mathcal{D}} Q\rho_b v_Q(x,r) \, d^2\r
  \nonumber\\
  &=& \frac{Qq}{2}
  \left[
    (N_b+\hx^2)\ln (N_b+\hx^2) -N_b
    -\hx^2\ln \hx^2
  \right]
  \,,
\end{eqnarray}
where we have chosen the arbitrary constant $\ell$ such that $\hl=1$.

%%%%%%%%%%%%%%%%%%%%%%%%%%%%%%%%%%%%%%%%%%%
\subsection{The special coupling $\Gamma=2$}
Since the one-component plasma on the disk is two-dimensional with
log-potential, one can use the special techniques developed for
two-dimensional Coulomb systems~\cite{J81, AJ81} and
random matrices~\cite{Mehta} to compute exactly the effective
interaction potential between the disk and the approaching charge, for
a special value of the Coulomb coupling $\Gamma=\beta q^2=2$.

Let $\r_i$ be the position of the $i$-th particle on the
disk, in polar coordinates $\r_i=(r_i,\varphi_i)$. It is convenient to
define $z_i=r_i e^{i\varphi_i}$ and $\hz_i=\hr_i e^{i\varphi_i}$.
The total potential energy of the system can be written, up to an irrelevant 
constant
\begin{equation}
  \label{eq:H}
  H=Qq \sum_{i=1}^{N} v_{Q}(x,r_i)+ V_{Qb}(x)
  +\frac{q^2}{2}\sum_{i=1}^{N} \hr_i^2 \,
  - q^2 \!\sum_{1\leq i< j \leq N} \ln|\hz_{i}-\hz_{j}| ,
%  +V_{bbp}
\end{equation}
%with
%\begin{equation}  \label{eq:Vbbp}
%  V_{bbp}=-\frac{q^2 N_b^2}{4} \left(
%    \ln N_b-\frac{1}{2}\right)
%  +\frac{q^2 N N_b}{2}
%  \left( \ln N_b -1\right)
%\,.
%\end{equation}
where the first two terms on the right hand side account for the test charge - mobile charge
and test charge - background
interactions respectively, while the last two terms are for the mobile charge - background and 
mobile charge - mobile charge energies.
When $\beta q^2=2$, up to a multiplicative constant, the Boltzmann factor reads
\begin{equation}
  \label{eq:ebetaH}
  e^{-\beta H}=
%  e^{-\beta V_{bbp}}
  e^{-\beta V_{Qb}(x)}
  \prod_{i=1}^{N}
  e^{-2 \frac{Q}{q} v_{Q}(x,r_i)-\hr_i^2}
  \prod_{1\leq i<j\leq N} |\hz_i-\hz_j|^2
  \,.
\end{equation}
The product $\prod_{1\leq i<j\leq N} (\hz_i-\hz_j)$ is a Vandermonde
determinant $\det(\hz_{i}^{j-1})$. Defining
\begin{equation}
  \label{eq:psi}
  \psi_{j}(\r)=
    e^{- \frac{Q}{q} v_{Q}(x,r)-\frac{\hr^2}{2}}
    \hz^{j}\,,
\end{equation}
the Boltzmann factor can be written as
\begin{equation}
  \label{eq:ebetaH2}
  e^{-\beta H}=
%%  e^{-\beta V_{bbp}}
  e^{-\beta V_{Qb}(x)}
  \left|
    \det\left(\psi_{j-1}(\r_i)\right)_{1\leq i,j \leq N}
  \right|^2
  \,.
\end{equation}
The functions $\psi_{j}$ are orthogonal
\begin{equation}
  \label{eq:ortho}
  \int_{\mathcal{D}}
  \overline{\psi_{j}(\r)}\psi_{k}(\r)
  \,d^2\r=0\,\qquad \text{if } j\neq k \,,
\end{equation}
with norm
\begin{equation}
  \label{eq:norm}
  \Vert\psi_j\Vert^2=\int_{\mathcal{D}} |\psi_j(\r)|^2\,d^2\r
  =\frac{1}{n_b}\int_{0}^{N_b}
  t^j (\hx^2+t)^{Q/q} e^{-t}\,dt
  \,.
\end{equation}
If $Q/q$ is a positive integer, this can be expressed in terms of
incomplete gamma functions $\gamma(k,N_b)=\int_0^{N_b} t^{k-1}
e^{-t}\,dt$. For instance, when $Q=q$,
\begin{equation}
  \label{eq:normgamma}
  \Vert\psi_j\Vert^2=\frac{1}{n_b}
  \left[
    \hx^2 \gamma(j+1,N_b)+ \gamma(j+2,N_b)
  \right]
  \,.
\end{equation}

The configurational canonical partition function is
\begin{eqnarray}
  Z&=&\frac{1}{N!} \int_{\mathcal{D}^N}
  e^{-\beta H} \,\prod_{i=1}^N d^2\r_i
  \nonumber\\
  &=&
  \frac{1}{N!}
  e^{-\beta V_{Qb}(x)}
  \int_{\mathcal{D}^N} 
  \left|
    \det\left(\psi_{j-1}(r_i)\right)_{1\leq i,j \leq N}
  \right|^2
  \,\prod_{i=1}^N d^2\r_i
  \,.
  \label{eq:Z}
\end{eqnarray}
If the determinant is explicitly expanded, and the integrals
performed, the result simplifies~\cite{AJ81,Mehta}, due to
the orthogonality of the functions $\psi_j$
\begin{equation}
  \label{eq:Zres}
  Z=e^{-\beta V_{Qb}(x)}
  \prod_{j=0}^{N-1} \Vert \psi_j \Vert^2
  \,.
\end{equation}
Up to an additive constant, the effective interaction potential,
$V_{\eff}(x)$, between the disk and the approaching charge $Q$, is
given by \cite{B00} $e^{-\beta V_{\eff}(x)} \propto Z$, and more specifically,
we choose 
\begin{equation}
  e^{-\beta V_{\eff}(x)} \, = \, \frac{Z}{Z_0}\,,
\end{equation}
where $Z_0$ is the $x$-independent partition function 
when $Q=0$. The above definition ensures that for $N=N_b$,  
$V_{\eff}(x) \to 0$ when $x\to \infty$.
On the other hand, for $N\neq N_b$, $V_{\eff}(x)$
diverges for $x\to \infty$, see below.
The physical meaning of $V_{\eff}$ is clear : $-\partial V_{\eff}(x) / \partial x$
provides the mean force experienced by $Q$, averaged over all possible fluctuations
of charge configurations on the disk. The function $V_{\eff}$ is precisely the free
energy of the system, for a given test charge - disk distance $x$.
Therefore,
\begin{eqnarray}
  \beta V_{\eff}(x)&=&
  \frac{Q}{q}
  \left[
    (N_b+\hx^2)\ln (N_b+\hx^2) -N_b
    -\hx^2\ln \hx^2
  \right]
  \nonumber\\
  &&
  -\sum_{j=1}^{N}
  \left[
  \ln \int_{0}^{N_b}
  t^{j-1} (\hx^2+t)^{Q/q} e^{-t}\,dt
  -\ln \gamma(j,N_b)
  \right]
  \,.
  \label{eq:Veff}
\end{eqnarray}
In the special case $Q=q$, we obtain
\begin{eqnarray}
  \beta V_{\eff}(x)&=&
    (N_b+\hx^2)\ln (N_b+\hx^2) -N_b
    -\hx^2\ln \hx^2
  \nonumber\\
  &&
  -\sum_{j=1}^{N}
  \ln \left[
    \hx^2 + \frac{\gamma(j+1,N_b)}{\gamma(j,N_b)}
    \right]
  \,.
  \label{eq:VeffQ=q}
\end{eqnarray}

The density profile $n(r)$ on the disk can also be obtained
explicitly~\cite{J81, Mehta}, and will be discussed in some
detail below
\begin{eqnarray}
  \label{eq:n}
  n(r)&=&
  \sum_{j=0}^{N-1} \frac{|\psi_{j}(\r)|^2}{\Vert \psi_j \Vert^2}
  \nonumber\\
  &=&
  n_b
  \sum_{j=0}^{N-1}
  \frac{\hr^{2j}(\hx^2+\hr^2)^{Q/q}\, e^{-\hr^2}}{\int_{0}^{N_b}
    t^{j} (\hx^2+t)^{Q/q}\, e^{-t}\,dt}\,.
\end{eqnarray}
It can be checked that the two situations where $Q=0$ and $x\to \infty$
are equivalent, since both decouple the test charge from those
on the disk.

%%%%%%%%%%%%%%%%%%%%%%%%%%%%%%%%%%%%%%%%%%%
\subsection{Arbitrary even coupling parameters}
\label{ssec:arb}

For couplings parameters $\Gamma=\beta q^2=2\gamma$, with $\gamma$ an integer,
the partition function of the system, and the effective potential, can
be computed for small enough number of particles $N$, by using
a method developed in~\cite{SPK94, TF99}, based on
techniques used in the study of the quantum Hall
effect~\cite{dFGIL94, D94, STW94}.  We 
provide here some details on the methods. 

Up to a multiplicative constant, the Boltzmann factor
of the system reads
\begin{equation}
  \label{eq:ebetaH-gamma}
  e^{-\beta H}=
%%  e^{-\beta V_{bbp}}
  e^{-\beta V_{Qb}(x)}
  \left|
    \det\left(\psi_{j-1}(r_i)\right)_{1\leq i,j \leq N}
  \right|^{2\gamma}
  \,.
\end{equation}
where, now, the orthogonal functions $\psi_{k}$ are
\begin{equation}
  \psi_{k}(\r)=[w(r)]^{1/2} \tilde{z}^k
\end{equation}
with
\begin{equation}
  w(r)=e^{-2\gamma\frac{Q}{q} v_{Q}(x,r)-\gamma \tilde{r}^2}
  \,.
\end{equation}
The key idea to compute the partition function is to expand
$[\det(\tilde{z}_{k}^{j-1})]^\gamma$ in terms of appropriate
orthogonal polynomials~\cite{TF99}. For $\gamma$ even, the expansion
is in terms of symmetric monomials, whereas for $\gamma$ odd, it is
expanded in terms of antisymmetric polynomials. The coefficients of
the expansion are conveniently indexed by a partition
$\mu=(\mu_1,\cdots, \mu_N)$ of $\gamma N(N-1)/2$, for example for
$\gamma$ even,
\begin{equation}
  [\det(\tilde{z}_{k}^{j-1})]^\gamma = 
  \sum_{\mu} c_{\mu}\,  \text{Sym}(z_1^{\mu_1}\ldots z_{N}^{\mu_N})
\end{equation}
with the symmetric monomial
\begin{equation}
\text{Sym}(z_1^{\mu_1}\ldots z_{N}^{\mu_N})
= \frac{1}{\prod_i m_i!}  
\sum_{\sigma \in S_{N}} z_{\sigma(1)}^{\mu_1} \ldots z_{\sigma(N)}^{\mu_N}
\end{equation}
where $S_{N}$ is the permutation group of $N$ elements and $m_{i}$ is
the multiplicity of the integer $i$ in the partition $\mu$. A similar
expression is used for $\gamma$ odd with antisymmetrized monomials.

Due to the orthogonality of the (anti)symmetric monomials, the
partition function is finally given also as an expansion similar to
the one of the power $\gamma$ of the Vandermonde determinant,
see~\cite{TF99} for details.  The final expression for the effective
potential is
\begin{equation}
  \label{eq:Veff-anygamma}
  \beta V_{\eff}(x)=\beta V_{Qb}(x)-\ln \frac{Z^{*}}{Z_0^{*}}
\end{equation}
with
\begin{equation}
Z^{*}=\sum_{\mu} \frac{c_{\mu}^2}{\prod_{i} m_i!}\, \prod_{k=1}^{N}
||\psi_{\mu_k}||^{2}  
\,,
\end{equation}
\begin{equation}
  ||\psi_{j}||^{2}=\int_{\cal D} w(r) r^{2j} d\r=
    \frac{1}{n_b} \int_{0}^{N_b} e^{-\gamma t}(\tilde{x}^2+t)^{\gamma
      Q/q} t^{j}\,dt
    \,,
\end{equation}
and $Z_0^{*}$ is $Z^{*}$ evaluated when $Q=0$.  In the case when
$\gamma$ is odd, the factor $\prod_{i} m_i!=1$, since due to the
antisymmetry, the admitted partitions $\mu$ do not have repeated
numbers.

This method can equivalently be formulated by transforming the
classical problem of the one-component plasma in a quantum problem of
a linear chain of interacting fermions, as explained in
Refs.~\cite{SP95, S04}. The starting point for this
method is to write the Vandermonde determinant as a Gaussian integral
over Grassmann variables. The final result is
again~(\ref{eq:Veff-anygamma}).

For the present work, we did some calculations up to $N=11$
particles. The coefficients $c_{\mu}$ needed for the numerical
calculations where kindly provided by L. \v{S}amaj for $\gamma=2$ up to
$N=10$ and for $\gamma=3$ up to $N=9$. For $\gamma=2$ and $N=11$, and
$\gamma=3$ and $N=10$ and $N=11$, we obtained the coefficients using
the algorithm recently proposed by Bernevig and
Regnault~\cite{BR09}, and their Jack polynomial
generator online code~\cite{jack}.
We now turn to the results obtained from the previous analysis, 
starting with the effective potential experienced by the test charge $Q$ 
at large distances from the 
disk.

%%%%%%%%%%%%%%%%%%%%%%%%%%%%%%%%%%%%%%%%%%%
%%%%%%%%%%%%%%%%%%%%%%%%%%%%%%%%%%%%%%%%%%%%%%%%%%%%%%%%%%%%%%%%%%%%%%%%%%5
\section{Long distance behavior}
\label{sec:long-distance}

\subsection{General results at arbitrary couplings}

As mentioned earlier, the effective interaction potential, also known as potential of mean
force, $V_{\eff}(x)$, has the property that $-\nabla V_{\eff}$ is the
mean force experienced by $Q$. It is interesting to introduce another
quantity, $V(x)$, the electric potential created by the average charge
density distribution $q (n(r)-n_b)$ at the position of the charge $Q$
\begin{equation}
  \label{eq:defV}
  V(x)=\int_{\cal D} q(n(r)-n_b) v_{Q}(x,r)\,d\r
\end{equation}
Because of the fluctuations and the fact that the presence of $Q$ at
position $x$ modifies the density on the disk, in general
$V_{\eff}(x)\neq Q V(x)$, only for a small infinitesimal charge $Q$
the equality holds. For arbitrary $Q$, a simple relation can be found
between the two, by noticing that the total potential energy of the
system~(\ref{eq:H}) depends linearly on $Q$, then
\begin{equation}
  \frac{\partial e^{-\beta H}}{\partial Q} =-\beta
  \int_{\cal D} q(\widehat{n}(r;\r_1,\ldots,\r_N)-n_b) v_{Q}(x,r)\,d\r \
  e^{-\beta H}
\end{equation}
where $\widehat{n}(r;\r_1,\ldots,\r_N)=\sum_{i=1}^{N}\delta(\r-\r_i)$ is
the microscopic density. Averaging this relation over all the
configurations of the ions on the disk, we find
\begin{equation}
  \label{eq:V-Veff}
 \frac{\partial V_{\eff}(x)}{\partial Q} = V(x)
 \,.
\end{equation}
At large distances from the disk, expanding $v_{Q}(x,r)$ for $r \ll
x$, one can obtain the multipolar expansion of the electric
potential $V$
\begin{equation}
V(x)=-q (N-N_b) \ln \hx -  q \frac{\mathbb{Q}_{2}}{2 x^2} + O(1/\hx^4)
\label{eq:multipol}
\end{equation}
where the relevant quadrupole moment $\mathbb{Q}_{2}$ results from
the second moment of the excess density $[n^{0}(r)-n_b]$
\begin{equation}
\mathbb{Q}_{2}\, = \, \int_\mathcal{D} r^2 [n^{0}(r)-n_b] \, d  \r 
\end{equation}
where $n^{0}(r)$ is the density when $x\to\infty$ (or
equivalently $Q=0$). Since we are computing the potential on the
$x$-axis, no dipolar contribution remains, while the logarithmic
monopole contribution stems from the global charge of the disk,
$q(N-N_b)$.  Since up to terms of higher power than $1/x^2$, 
Eq.~(\ref{eq:multipol}) shows that $V(x)$ does not depend on $Q$,
integrating Eq.~(\ref{eq:V-Veff}) one finds that Eq.~(\ref{eq:multipol})
also gives the large $x$ expansion of $V_{\eff}$ (multiplied by $Q$).

We also mention here a sum rule that turns out interesting for the
following discussion. The quadrupolar moment $\mathbb{Q}_{2}$
can be shown to be related to the mobile particle density 
at contact through \cite{CFG80,TF99}
\begin{equation}
\Gamma \,\frac{n_b}{2 R^2} \,\mathbb{Q}_{2} \,=\, 
n_b \left(1-\frac{\Gamma}{4} \right) - n^{0}(R)
\label{eq:sumquad}
\end{equation}
for a neutral disk.

For the next order of the expansion, in $1/x^4$, the situation is more
involved. The expansion of $V(x)$ cannot be obtained only from the
next multipole $\mathbb{Q}_4=\int_{\cal D} r^4 (n^0(r)-n_b)\,d\r$,
since the density $n(r)$ itself depends also on $x$ and one needs to
take into account the next-to-leading order of the large-$x$
expansion of $n(r)$ to compute properly the expansion of $V(x)$ up to
order $1/x^4$. This next-to-leading order is of order $1/x^2$ and it
is proportional to $Q$ as it can be checked by expanding $e^{-\beta
  H}$ for large $x$. It has the form
\begin{equation}
  n(r)=n^{0}(r)+\frac{\beta Q q d_2(r)}{x^2} + O(1/\hx^4)
\end{equation}
where $d_2(r)$ is a function only of $r$ and $\beta q^2$.
Using this expansion one can obtain from Eq.~(\ref{eq:defV}) the
expansion of $V(x)$ up to the order $1/x^4$
\begin{equation}
V(x)=-q (N-N_b) \ln \hx -  q \frac{\mathbb{Q}_{2}}{2 x^2} 
+ q\left (\frac{\mathbb{Q}_4}{4} - \frac{\beta qQ}{2} \int_{\cal D}
r^2  d_2(r) \,d\r \right)
\frac{1}{x^4} + O(1/x^6)
\,.
\label{eq:multipol4}
\end{equation}
Then integrating with respect to $Q$ one finds
\begin{equation}
  V_{\eff}(x)=-q Q (N-N_b) \ln \hx -  q Q \frac{\mathbb{Q}_{2}}{2 x^2} 
+ qQ \left (\frac{\mathbb{Q}_4}{4} - \frac{\beta qQ}{4} \int_{\cal D}
r^2  d_2(r) \,d\r \right)
\frac{1}{x^4} + O(1/x^6)
\,.
\end{equation}
The term involving the second moment of $d_2$ differs by a factor
$Q/2$ between $V(x)$ and $V_{\eff}(x)$. In the following section we
illustrate these considerations on the explicit results obtained when
$\Gamma=\beta q^2=2$.

\subsection{Results at $\Gamma=2$}

Unless otherwise specified, the results reported correspond to
$\Gamma=2$.  For $Q$ arbitrary, the long distance behavior of the
effective interaction~(\ref{eq:Veff}), for $N$ and $N_b$ fixed,
$\hx^2\gg N$ and $\hx^2\gg N_b$ is
\begin{eqnarray}
  \label{eq:Veffasympttmp}
  V_{\eff}(x)&=&-Q q (N-N_b) \ln \hx 
  + \frac{Q q}{2}\frac{1}{\hx^2} 
  \left[\frac{N_b^2}{2}-\sum_{j=1}^{N}
    \frac{\gamma(j+1,N_b)}{\gamma(j,N_b)}\right]
  \\
  &&-\frac{Q}{4 \hx^4} \sum_{j=1}^N \left[
  \left(Q-q\right)\frac{\gamma(j+2,N_b)}{\gamma(j,N_b)}
  - Q\frac{\gamma(j+1,N_b)^2}{\gamma(j,N_b)^2}
  \right]
  +O(1/\hx^6)\, ,
  \nonumber
\end{eqnarray}
the structure of which deserves some comments.  Up to order $1/x^2$,
such a series has the form of a multipolar expansion, in agreement
with the discussion from the previous section. Indeed, the coefficient
of $1/\hx^2$ is precisely $-qQ\mathbb{Q}_{2}/2$ as it can be checked
by computing the second moment of the excess density from the explicit
expression~(\ref{eq:n}) when $Q=0$. Eq.~(\ref{eq:Veffasympttmp}) can be
compared to the large-$x$ expansion of the electric potential
\begin{eqnarray}
  \label{eq:Velecasympt}
  V(x)&=&- q (N-N_b) \ln \hx 
  + \frac{q}{2}\frac{1}{\hx^2} 
  \left[\frac{N_b^2}{2}-\sum_{j=1}^{N}
    \frac{\gamma(j+1,N_b)}{\gamma(j,N_b)}\right]
  \\
  &&-\frac{1}{4 \hx^4} \sum_{j=1}^N \left[
  \left(2Q-q\right)\frac{\gamma(j+2,N_b)}{\gamma(j,N_b)}
  - \frac{\gamma(j+1,N_b)^2}{\gamma(j,N_b)^2}
  \right]
  +O(1/\hx^6)\, ,
  \nonumber
\end{eqnarray}
where one can explicitly check that $\partial_{Q} V_{\eff}=V$.

Let us discuss further the expansion of $V_{\eff}$ up to order
$1/x^2$. Using the properties of the incomplete gamma function, that
allow us to write the coefficient of the $1/\hx^2$ term appearing in
Eq. (\ref{eq:Veffasympttmp}) as
\begin{eqnarray}
  \frac{N_b^2}{2}-\sum_{j=1}^{N}
  \frac{\gamma(j+1,N_b)}{\gamma(j,N_b)}
  &=&
  \frac{N_b^2-N^2}{2}-\frac{N}{2}+\sum_{j=1}^{N}
  \frac{e^{-N_b}N_{b}^{j}}{\gamma(j,N_b)}
  \nonumber\\
  &=&
  \frac{N_b^2-N^2}{2}-\frac{N}{2}+\frac{N_b n^{0}(R)}{n_b}
  \,,
  \label{eq:x2}
\end{eqnarray}
where $n^{0}(R)$ is the density of particles at the edge of the disk
in the absence of the charge $Q$, i.e.~Eq.~(\ref{eq:n}) with $Q$=0 at
$\hr=\hR=\sqrt{N_b}$. Thus, 
\begin{equation}
  \label{eq:Veffasympt}
  V_{\eff}(x)=-Q q (N-N_b) \ln \hx 
  + \frac{Q q}{2}\frac{1}{\hx^2} 
  \left[\frac{N_b^2-N^2}{2}-\frac{N}{2}+\frac{N_b n^{(0)}(R)}{n_b}
      \right]
  +O(1/\hx^4)\,.
\end{equation}

\begin{figure}
  \centering
  \includegraphics[width=\GraphicsWidth]{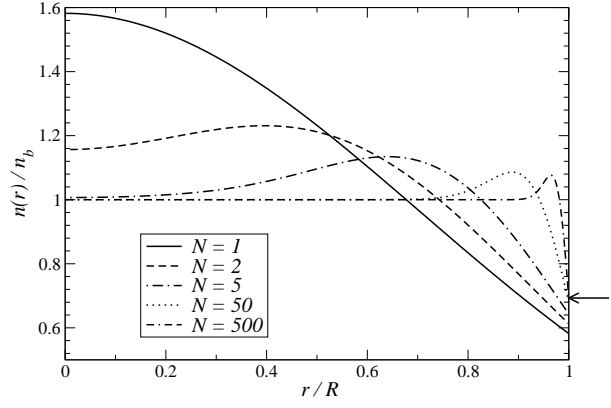}
  \vspace{5mm}
  \caption{\label{fig:ndens.xstar.infty} Reduced charge density
    profile in the disk, for different neutral situations ($N=N_b$),
    and $x\to \infty$. The arrow on the right hand side indicates the
    limiting value $\ln 2 \simeq 0.693$ that is reached in the large
    $N$ limit. The total charge density profile is $q[n(r)-n_b]$. For
    large $N$, it thus vanishes except in a small region of linear
    size $1/\sqrt{N_b}$ in the vicinity of the boundary $r=R$.  }
\end{figure}

For a neutral disk, $N=N_b$, and
\begin{equation}
  \label{eq:Veffasymptneutral}
  V_{\eff}(x)=
  \frac{Q q N_b}{2}\frac{1}{\hx^2} 
  \left[\frac{n^{(0)}(R)}{n_b}
    -\frac{1}{2}
      \right]
  +O(1/\hx^4)\, .
\end{equation}
This result is an explicit check, at $\Gamma=2$, of the multipolar
expansion~(\ref{eq:multipol}) combined with the sum rule
(\ref{eq:sumquad}). One notice that the $V_{\eff}$ is repulsive for
$Qq >0$. This can be understood as follows.  When $\hx\to\infty$, the
charge density profile $n(r)$ inside the disk is the same one as for
a disk alone (without the approaching charge $Q$), found in
Refs.~\cite{J81,J82}. The density $n(r)$ is equal to the background
density $n_b$ in the bulk of the disk (local neutrality in the
bulk). Close to the boundary, it raises above the background density,
then falls below it~\cite{J82}, see
figure~\ref{fig:ndens.xstar.infty}.  Therefore, there are two
concentric layers of charges close to the edge: the inner layer bears
a net charge which is of the same sign as $q$ [i.e.~$n(r)>n_b$],
while the outer one is opposite, by electroneutrality. This ensures
that $\mathbb{Q}_{2}$ is generically negative, so that the
quadrupolar term yields an effective interaction (disk-test charge)
that is of the same sign as $Q q$, i.e.~repulsive for $Q q>0$.  

When $N\to\infty$, more explicit results can be obtained. In this limit,
the density at the edge of the disk takes a simple
form~\cite{J82}, $n^{(0)}(R)=n_b \ln 2$, so that
\begin{equation}
  \label{eq:VeffasymptneutralNinf}
  V_{\eff}(x)=  
  \frac{Q q N_b}{2}\frac{1}{\hx^2} 
  \left(\ln 2
    -\frac{1}{2}
      \right)
  +O(1/\hx^4)\,
= \frac{Q q R^2}{2}\frac{1}{x^2} 
  \left(\ln 2
    -\frac{1}{2}
      \right)
  +O(1/x^4) ,
\end{equation}
with $\ln 2 -\frac{1}{2}\simeq 0.19 >0$, which is consistent with the
generic discussion above on the negative sign of
$\mathbb{Q}_{2}$. Again, the effective potential is attractive at
large distances for $Qq<0$, repulsive for $Qq>0$, for a neutral disk.

The quadrupolar route allows us to obtain results for strongly coupled
systems (large $\Gamma$), making use of the sum rule
(\ref{eq:sumquad}).  We note in passing that this general result is
compatible with the value $\mathbb{Q}_{2} = - q R^2 (\log 2- 1/2)$
that holds at $\Gamma=2$ when the number of mobile charges on the disk
becomes large, see Eqs. (\ref{eq:VeffasymptneutralNinf}) and
(\ref{eq:multipol}).  When $\Gamma$ itself turns large, the system
crystallizes, but the sum rule~(\ref{eq:sumquad}) remains
valid, provided $n(R)$ is replaced by the average of the contact
density over the perimeter of the disk~\cite{CFG80,TF99}. It is
physically reasonable to suppose that the average of $n(R)/n_b$
remains bounded in this limit. Actually, for the three-dimensional
analogue of this model, with $1/r$ interaction, this is the
case~\cite{HA91}. Then Eq.~(\ref{eq:sumquad}) becomes $\mathbb{Q}_{2}
\sim -R^2/2$. We consequently have
\begin{equation}
  V_{\eff}(x) \underset{\Gamma \to \infty}{\sim}  \frac{Q q R^2}{4}\frac{1}{x^2}  
  +O(1/x^4) ,
\end{equation}
which is again repulsive for $Q q >0$, attractive otherwise.
It can be mentioned here that the scaling result $\mathbb{Q}_{2} \propto - R^2$ 
is readily recovered by the two concentric layers simplified viewpoint. 
For large $N$, there exists a outermost corona void of charges:
particles are depleted there, as they are in the plum pudding model,
see e.g. \cite{HA91,BP94,CCRT11}, the width of which is given by the
typical distance between particles $\delta \propto R/\sqrt{N}$
(at $\Gamma=2$, $\delta$ is already the typical distance between
the density maximum and the disk radius that can be seen in Fig.
\ref{fig:ndens.xstar.infty} for large $N$). The charge in this corona
is given by $-q  n_b R \delta$, which contributes a quantity 
$- R \, n_b \, \delta \, R^2$ to the quadrupole moment. This charge is
compensated by an oppositely charged ring, located at $R-\delta$,
which contributes a quantity $R \, \delta \, n_b (R-\delta)^2$ to 
$\mathbb{Q}_{2}$. Summing both contributions, assuming that the
particles and background with $r<R-\delta$ do not contribute to $\mathbb{Q}_{2}$, 
and remembering that 
$\delta \ll R$, we arrive at 
$\mathbb{Q}_{2} \propto - n_b R^2 \delta^2 \simeq - R^2 $.
A very similar argument holds at $\Gamma=2$, since
the two corona approach is also valid.

%%%%%%%%%%%%%%%%%%%%%%%%%%%%%%%%%%%%%%%%%%%%%%%%%%%%%%%%%%%%%%%%%%%%%%%%%%
\section{Short scale features}
\label{sec:short}

We now turn to the study of the phenomenology at shorter distances, which
is different depending on whether the particle approaching
the disk has a charge of the same sign of the mobile particles on the
disk ($Q/q>0$), or a charge of opposite sign. In addition, the cases
of globally neutral or charged disks should be treated separately, and
the different cases are ruled by different sorts of
polarization effects.

\subsection{Case $Q/q>0$}
\label{sec:Qpos}

%%%%%%%%%%%%%%%%%%%%%%%%%%%%%%%%%%%%%%
\subsubsection{Neutral disk}

We consider a globally neutral disk $N=N_b$. We study in this section
if it is possible to overcharge this object, by approaching a particle
that has a charge $Q$ with the same sign of the mobile counterions on
the disk. At large distances, we know from~(\ref{eq:Veffasympt}) that
the interaction is repulsive. We anticipate that this behavior
changes when the charge $Q$ is close enough to the disk,
since the intruder $Q$ should then create a correlation hole, pushing 
mobile charges closer to the boundary $r=R$, and thereby gaining
Coulombic energy from hole opened. This is the mechanism behind
charge inversion in colloidal systems, that has been reviewed
for situations of strong coupling 
in Ref.~\cite{GNS02}.

\begin{figure}[htb]
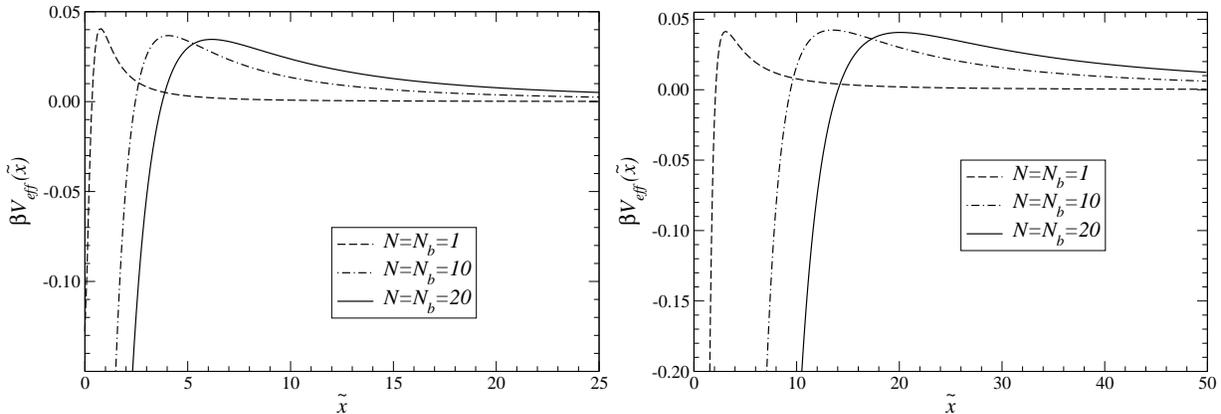

  \null\vskip 5mm
  \centering
  \includegraphics[width=\GraphicsWidth]{veffQ=1}\hspace{1mm}\includegraphics[width=\GraphicsWidth]{veffQ=10}
  \null\vskip 5mm
  \caption{\label{fig:veffQ1-N=Nf} 
    The effective interaction between the disk and an approaching ion
    with charge $Q=q$ (left graph) or $Q=10 q$ (right graph). 
    The disk is globally neutral with $N_b=N$ ions
    of charge $q$. The distance $x$ is expressed in reduced units ($\hx$),
    in which the disk radius reads $\sqrt{N_b}$, and therefore takes different
    values for the three curves shown.
  }
\end{figure}

The short-distance behavior of the effective potential~(\ref{eq:Veff})
is, when $Q/q>0$,
\begin{eqnarray}
  \beta V_{\eff}(x)&=&
  \frac{Q}{q} \left[ N_b\ln N_b-N_b
    +\hx^2 \left(1+\ln\frac{N_b}{\hx^2}\right)
    \right]
    -\sum_{j=1}^{N}
    \ln\frac{\gamma\left(j+\frac{Q}{q},N_b\right)
    }{\gamma\left(j,N_b\right)}
    \nonumber\\
    &&-
    \frac{Q}{q}\hx^2\sum_{j=1}^{N}
  \frac{\gamma\left(j+\frac{Q}{q}-1,N_b\right)
  }{\gamma\left(j+\frac{Q}{q},N_b\right)}
  +O\left(\hx^4,\hx^{2(1+Q/q)}\right)
  \,,
  \label{eq:veff-shortQpos}
\end{eqnarray}
which is clearly an increasing function of $x$ when $\hx\ll1$. Therefore, at
close distance from the disk, the interaction turns out to be
attractive. This can be observed in figure~\ref{fig:veffQ1-N=Nf}, where
the effective potential indeed increases at
short distances, reaches a maximum and then decreases upon increasing the distance
between the test particle and the disk. Note that at
$x=0$, the effective potential takes a finite value
\begin{equation}
  \label{eq:betaVeff0}
  \beta V_{\eff}(0)=\frac{Q}{q} \left(N_{b}\ln N_{b}-N_{b}\right)
  -\sum_{j=1}^{N}
  \ln\frac{\gamma(j+\frac{Q}{q},N_{b})}{\gamma(j,N_{b})}
  \,,
\end{equation}
although it is not shown in all figures.

Let $x^{*}$ be the distance at which the interaction potential reaches
its maximum, and $\hx^{*}=\sqrt{\pi n_b} x^{*}$.  $x^{*}$ is the
minimum distance that one has to approach the charged particle in
order to overcome the natural repulsion of the disk. The corresponding
(free) energy cost to overcharge the disk is
given by $V^{\dagger}=V_{\eff}(x^{*})$, and
$V^{*}=V^{\dagger}-V_{\eff}(0)$ is the binding energy, i.e. the
necessary energy to unbind the charged particle from the disk, once it
has been overcharged. More generally, $V^{*}$ can be defined as the energy to 
overcome to peel off an ion from the disk. In all the present
discussion, the energy costs alluded to correspond to the work
an external operator holding the intruder should perform,
and this equals the corresponding free energy variation of the system as 
a whole. 

\begin{figure}[htb]
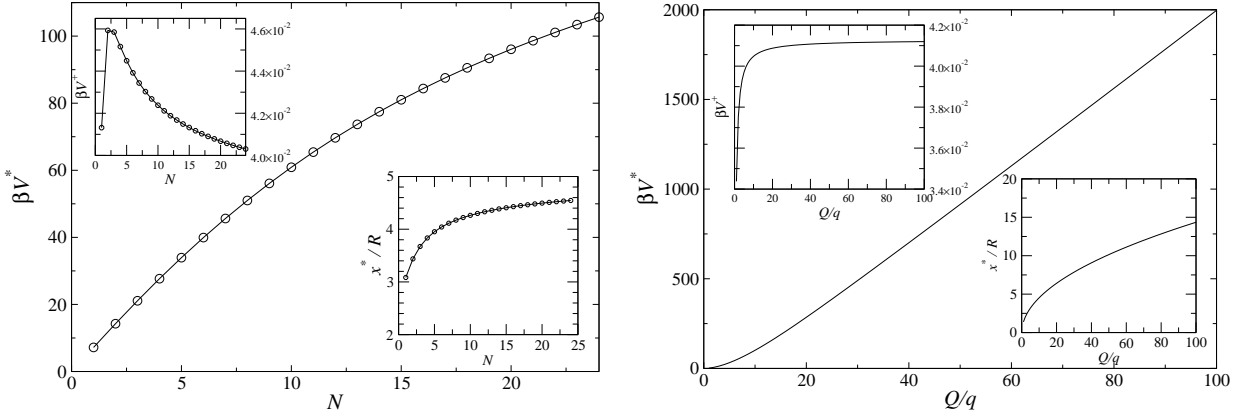

  \null\vskip 6mm
  \centering
  \includegraphics[width=\GraphicsWidth]{Vdagger-Vstar-xstar-fnct-N=Nb-Q10} 
  \hspace{1mm}
  \includegraphics[width=\GraphicsWidth]{Vdagger-Vstar-xstar-fnct-Q-N=Nb=22}
  \vspace{5mm}
  \caption{\label{fig:Vdagger-Vstar-xstar-fnct-N=Nb-Q10} Left: The binding
    energy $V^{*}$, the energy cost $V^{\dagger}$ and the distance
    $x^{*}$ to overcharge the globally neutral disk with an additional
    particle of charge $Q=10q$, as a function of the number of particles
    $N=N_b$ on the disk.
    Right: same quantities as a function of intruder charge $Q$, for $N=N_{b}=22$.}
\end{figure}

Figure~\ref{fig:Vdagger-Vstar-xstar-fnct-N=Nb-Q10} shows how $x^{*}/R$,
$V^{\dagger}$ and $V^{*}$ depend on $N$ for fixed $Q/q$, and on $Q/q$
at fixed $N$, respectively. First of all, it appears that 
the binding energy $V^{*}$ is several orders of magnitude
larger than the energy cost $V^{\dagger}$. This means that 
$V^* \simeq |V_{\eff}(0)|$. Second, the threshold distance
$x^*$ scales like $R$, when $N$ becomes large
enough, a fact that is masked in Fig. \ref{fig:veffQ1-N=Nf}
by the choice of units made (tilde variables, for which 
$\widetilde R = \sqrt{N_b}$). More precisely, from the numerical data of
Fig.~\ref{fig:Vdagger-Vstar-xstar-fnct-N=Nb-Q10}, we explored how
$x^{*}/R$ depends on the charge $Q$. We found, numerically, the
approximate relation
\begin{equation}
  \frac{x^{*}}{R\ \ }=a \sqrt{\frac{Q}{q}} +O(1/\sqrt{N})
\end{equation}
with $a=1.5+O(1/\sqrt{N})$. 
The fact that $a$ is of order one means that the effect of
effective attractions holds up to rather large distances,
on the order of the disk radius.
Another feature visible on Fig. \ref{fig:Vdagger-Vstar-xstar-fnct-N=Nb-Q10} is that 
the energy cost increases with $Q/q$, but quickly saturates to a
finite value $V_{\eff}^{\text{sat}}$. On the other hand
the binding energy increases as the charge $Q$ increases, as expected,
but it also increases with the number of mobile ions on the disk
$N$. For $N\to\infty$, using Stirling formula for the incomplete gamma functions in Eq.~(\ref{eq:betaVeff0}), one can obtain the analytical behavior of $V_{\eff}(0)$, and therefore the one of the binding energy, remembering that $V^{*}\simeq |V_{\eff}(0)|$. We find
\begin{equation}
  \label{eq:betaV0Ninfty}
  \beta V_{\eff}(0)= -\frac{1}{2} \left(\frac{Q}{q}\right)^2
  \ln N + O(1) \,.
\end{equation}

\begin{figure}
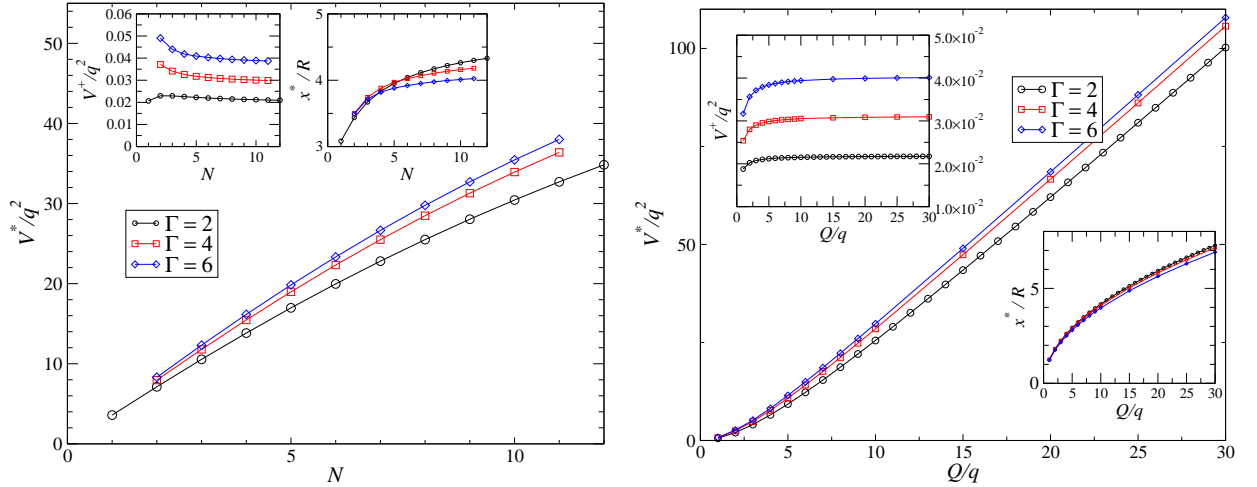

  \centering
  \includegraphics[width=\GraphicsWidth]{Vdagger-Vstar-xstar-fnct-N=Nb-Q10-allgamma}
  \hspace{1mm}
  \includegraphics[width=\GraphicsWidth]{Vdagger-Vstar-xstar-fnct-Q-N=Nb=8-allgamma}
  \vspace{5mm}
  \caption{\label{fig:Vdagger-Vstar-xstar-fnct-N=Nb-Q10-allgamma} Left: The binding
    energy $V^{*}$, the energy cost $V^{\dagger}$ and the distance
    $x^{*}$ to overcharge the globally neutral disk with an additional
    particle of charge $Q=10q$ as a function of the number of particles
    $N=N_b$ on the disk, for different values of the Coulombic
    coupling $\Gamma=\beta q^2$. Right : same as a function of $Q$, for a
    globally neutral disk with $N=N_{b}=8$.
    }
\end{figure}

Figure~\ref{fig:Vdagger-Vstar-xstar-fnct-N=Nb-Q10-allgamma} shows how the
previous quantities behave under different couplings
($\Gamma=\beta q^2=2, 4, 6$). The qualitative features appear to be robust:
The behavior is similar to
the one when $\Gamma=2$, with changes in the numerical values of
$x^{*}$, $V^{\dagger}$ and $V^{*}$. As $\Gamma$ increases,
$V^{\dagger}$ and $V^{*}$ increase, and $x^{*}$ decreases slightly:
the test ion has to come closer to the disk, and requires more
energy to overcome the long distance repulsion. Once it overcharges the disk
it is more energetically bounded to it.

\subsubsection{Charged disk}

If $N<N_b$, the disk has a charge $q(N-N_b)$ of opposite sign to that of the
approaching ion $Q$. In this case the effective potential is attractive
at all distances.
Let us consider the more interesting case where the disk is already
overcharged, with a net charge of the same sign as $Q$,
i.e.~$N>N_b$. The question is to study the effective potential
profile, and the distance range where it corresponds to an effective
attraction.

\begin{figure}
  \null\vskip 6mm
  \centering
  \includegraphics[width=\GraphicsWidth]{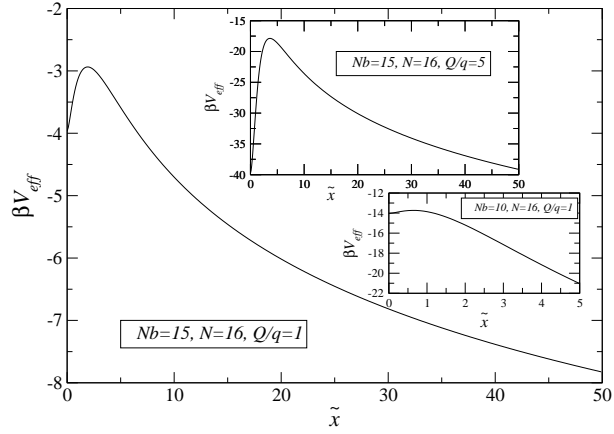}
  \vspace{5mm}
  \caption{\label{fig:veffNb-lt-N} 
    The effective potential between the charged disk and the
    approaching ion in cases where $Q/q>0$ and where the large distance
    behaviour is repulsive (i.e. $N>N_b$). The main inset is for 
    $N_b=15,N=16, Q/q=5$ while the smaller inset is for
    $N_b=10,N=16, Q/q=1$.
  }
  \vspace{5mm}
\end{figure}

\begin{figure}
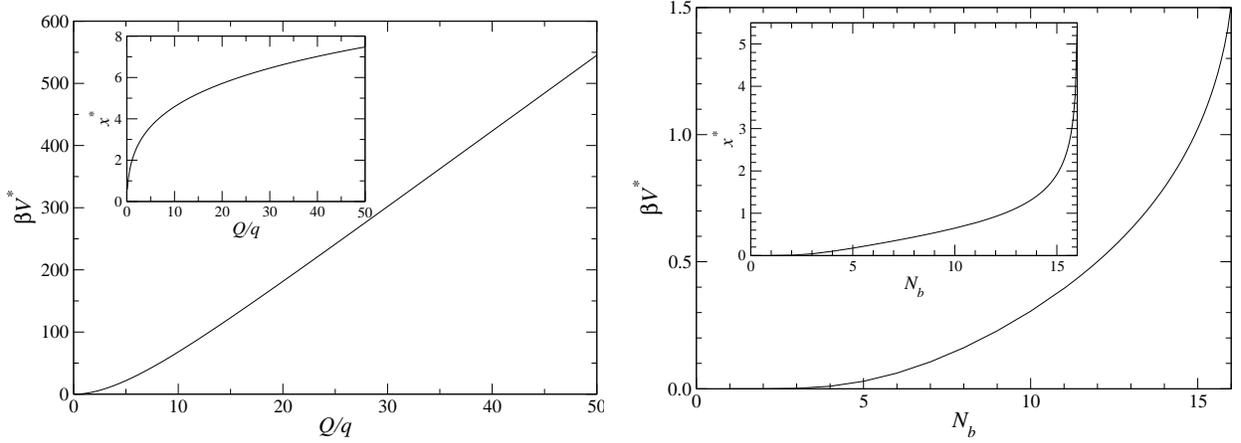

  \centering
  \vspace{6mm}
  \includegraphics[width=\GraphicsWidth]{Q-xstar-vstar-Nb=15-N=16}
  \hspace{1mm}
  \includegraphics[width=\GraphicsWidth]{Nb-xstar-vstar-N=16-Q=1}
  \vspace{5mm}
  \caption{\label{fig:Q-xstar-vstar-Nb=15-N=16} Left: The binding energy
    $V^{*}$ and the distance $x^{*}$ to overcharge the charged disk
    with $N_{b}=15$ and $N=16$ particles (global charge $q$) with an
    additional charged particle with charge $Q$ as a function of $Q/q$.
    Right: Same, as a function of $N_b$, for $N=16$ and $Q=q$.}
\end{figure}

At large distances, the effective interaction between the ion and the
disk is repulsive, and diverges as $-Qq(N-N_b)\ln \hx$. However, from
Eq.~(\ref{eq:veff-shortQpos}), we find that, at  short distances,
the effective potential becomes attractive. Therefore, as in
the previous situation, there exists a distance $x^{*}$ below which the test charge
will be attracted, which results in a further charge inversion of the disk. 
Figure~\ref{fig:veffNb-lt-N} shows the effective potential in this
situation for different charges $Q$ and charges of the disk
$q(N-N_b)$. Here, one can also define a binding energy
$V^{*}=V_{\eff}(x^{*})-V_{\eff}(0)$, necessary to pull out the ion from
the disk once it has been ``adsorbed''. 
Figure~\ref{fig:Q-xstar-vstar-Nb=15-N=16} shows how $x^{*}$ and $V^{*}$ depend
on the charge $Q$ of the ion and on the charge of the disk,
respectively.
It is also useful to
emphasize that when the disk complex is not neutral, one cannot
define the energy barrier $V^{\dagger}$. Indeed, this quantity 
was defined in the neutral case as the barrier to overcome
to approach the test charge from $x=\infty$, down to the distance
where attraction sets in. When $N\neq N_b$, the large distance 
effective potential diverges as $(N-N_b) \log x$, which precludes
the definition of $V^{\dagger}=V_{\eff}(x^*)-V_{\eff}(\infty)$. This 
feature is absent in three dimensions, where charges interact 
though a $1/r$ potential (hence the possible definition of $V^{\dagger}$
also for non neutral complexes). As a consequence, the study of overcharging
is somewhat less rich in the present case than for three dimensional
systems and overcharging is, with a log potential, necessarily a phenomenon  
of small amplitude (if not infinitesimal) : another way to rephrase
previous remarks is that $V^{\dagger}$ diverges as soon as  $N\neq N_b$.

\begin{figure}
  \centering
  \vspace{6mm}
  \includegraphics[width=\GraphicsWidth]{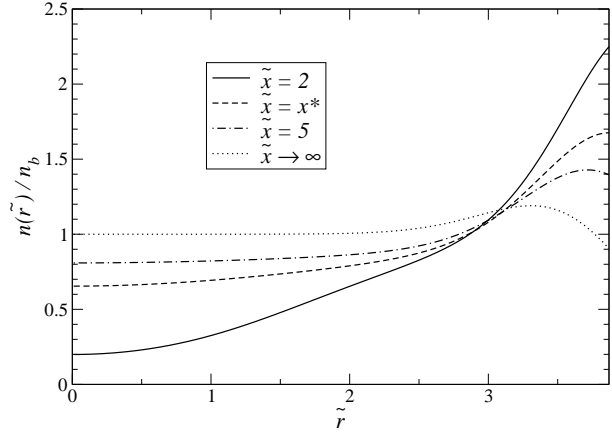}
  \vspace{5mm}
  \caption{\label{fig:ndens-Nb=15-N=16-Q=5-x=2-xstar-5-infty} The
    density profile of mobile particles in the disk, with $N=16$,
    $N_b=15$, charge of the disk equal to $q$ and the approaching ion
    has charge $Q=5q$. Notice that as the ion approaches the disk, the
    charge density in the center of the disk becomes negative. This
    results in the effective attraction at short-distances $x$ of the
    disk and the ion.  }
\end{figure}

To understand the mechanism behind the attraction at short distances,
it is instructive to study the density distribution of particles in
$\mathcal{D}$, as $Q$ approaches the disk.
Figure~\ref{fig:ndens-Nb=15-N=16-Q=5-x=2-xstar-5-infty} shows the
density profile for different distances $x$. It can be seen 
that the correlation hole alluded to earlier is increasingly
marked, when $x$ becomes smaller: the mobile charges $q$ of the disk
feel the repulsion due to the charge $Q$, and move towards the
edge of the disk. This results in a local negative charge density in
the center of the disk, which is finally responsible for the
attractive interaction between the disk and the intruder charge $Q$.

\subsection{Case $Q/q<0$}

We now turn to the case where the test particle and mobile ions 
on the disk have charges of opposite signs.

\subsubsection{Neutral disk}

\begin{figure}[htb]
  \null\vskip 6mm
  \centering
  \includegraphics[width=\GraphicsWidth]{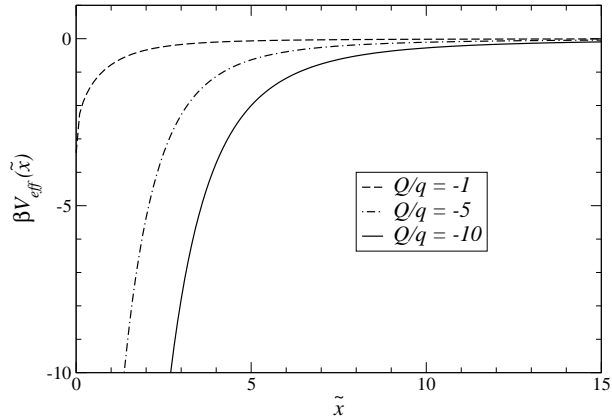}
  \vspace{5mm}
  \caption{\label{fig:veffQnegN=10} The effective potential
    between the globally neutral disk and the approaching ion, in the case where
    the charge of the test ion and those of the mobile particles on
    the disk have opposite signs: $Q/q<0$. Here, the disk bears $N=10$
    particles with charge $q$. }
  \vspace{5mm}
\end{figure}

For a globally neutral disk ($N=N_b$), we know 
from Eq.~(\ref{eq:Veffasymptneutral}) and the analysis of
section~\ref{sec:long-distance}
that the effective
potential is attractive at large-distances. This behavior remains at
short-distances, as illustrated in
figure~\ref{fig:veffQnegN=10}. Therefore, the neutral disk has a
natural tendency to overcharge.

\subsubsection{Charged disk}

If $N>N_b$, the disk has a charge $q(N-N_b)$ of opposite sign as the
approaching ion $Q$. In this case the effective potential is always
attractive, a situation that is not of particular interest.
We concentrate instead on the case where the disk has a net charge 
of the same sign as $Q$,
i.e.~$N_b>N$. Due to this excess charge, the effective potential with a
charge $Q$ of the same sign as the disk is expected to be repulsive at
large distances, see equation~(\ref{eq:Veffasympt}). However, as we
shall see below,  there is here also a change in the behavior
of the effective interaction at short distances, where the force between the disk and the
particle becomes attractive. 

The situation seems at first sight similar to the case studied in 
section~\ref{sec:Qpos}.  There are some notable
differences though. If $Q/q\leq -1$, the effective potential diverges when
$\hx\to0$. Indeed, for $Q/q\leq -1$, equation~(\ref{eq:veff-shortQpos}) is
no longer valid. The dominant contribution is given
by the term $j=1$ in the sum~(\ref{eq:Veff}). Explicitly, it yields, for $Q/q<-1$,
\begin{equation}
  V_{\eff}(x)\sim -q(Q+q) \ln \hx\,,\qquad x\to0
  \label{eq:fin1}
\end{equation}
and
\begin{equation}
  V_{\eff}(x)\sim -\frac{q^2}{2}\ln \left(\ln \frac{1}{\hx}\right)\,,\qquad x\to0\,,
  \label{eq:fin2}
\end{equation}
if $Q=-q$. In both cases, the small $x$ behaviour is attractive. We
note that the divergence of $V_{\eff}(x)$ for $x\to 0$ stems from the fact that 
the Boltzmann weight $\exp(\beta Q q\log r)=1/r^{-\Gamma Q/q}$
is non integrable at $r=0$ whenever $Q/q < -2/\Gamma$.
As announced earlier, the
effective potential is repulsive at large distances and attractive at
short distances. We can thus again define the distance $x^{*}$ at which the
potential becomes attractive if we approach the ion below
$x^{*}$. However, in the present case, the binding
energy $V^{*}$ is infinite because $\lim_{x\to0} V_{\eff}(x)=-\infty$.
Likewise, we cannot define the energy barrier $V^\dagger$ since
either attraction applies at all distances (neutral complex on the disk), or the effective potential
diverges at infinity (charged case).

\begin{figure}[htb]
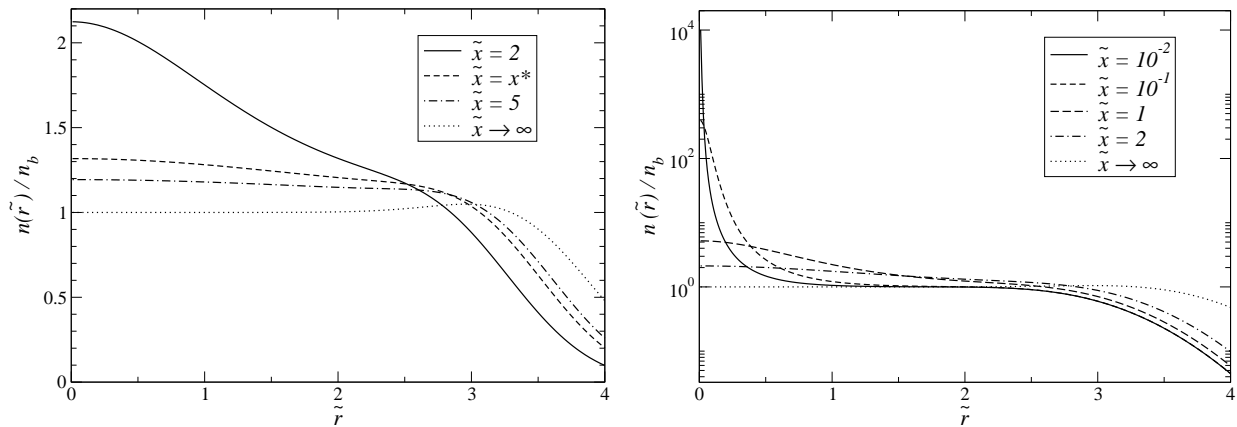

  \null\vskip 6mm
  \centering
  \includegraphics[width=\GraphicsWidth]{ndens-Nb=16-N=15-Q=-5-x=2-xstar-5-infty} \hspace{1mm}
\includegraphics[width=\GraphicsWidth]{logndens-Nb=16-N=15-Q=-5-x=0p01-0p1-1-2-infty}  
  \null\vspace{5mm}
  \caption{\label{fig:ndens-Nb=16-N=15-Q=-5-x=2-xstar-5-infty} Density
    profile of counterions in the disk, in the case $Q=-5q$, with $N=15$
    and $N_b=16$. The net charge of the disk is then equal to $-q$.   }
  \vspace{6mm}
\end{figure}
The mechanism behind the attraction between these like-charged objects
at short distance is now due to an accumulation of mobile charges near
the center of the disk. This can be seen in
Fig.~\ref{fig:ndens-Nb=16-N=15-Q=-5-x=2-xstar-5-infty}, which shows
the density profile of the mobile charges $q$, ie.~the counterions. As
the charge $Q$ approaches, it strongly attracts the counterions to the
disk center, and this results in an effective attraction, the precise
form of which is non trivial.  As seen on
Fig.~\ref{fig:ndens-Nb=16-N=15-Q=-5-x=2-xstar-5-infty}, for small
$\hx$, there is a large density of counterions close to the center of
the disk, but which is concentrated over a disk of radius of order 1 in $\hr$ units. The precise analytical behavior of the charge density can be extracted from Eq.~(\ref{eq:n}), for $\hx\ll 1$, taking into account that for $Q/q<-1$ the first term of the sum is the dominant one
\begin{equation}
  \label{eq:dens-x0}
  n(r)\sim - n_b\, e^{-\hr^2} \left(\frac{Q}{q}+1\right) 
  \left(1+\frac{\hr^2}{\hx^2}\right) \, 
  \frac{1}{\hx^2}
  \,.
\end{equation}
The total integrated charge of this counterion cloud close to the
center plus the background turns out to be equal to $-Q$, as one might
expect. However, the behaviour of the effective potential encoded in
Eqs. (\ref{eq:fin1}) and (\ref{eq:fin2}) exhibits a different
attraction than the bare Coulombic form $-Q^2 \log \widetilde x$, that
would be obtained assuming the attracted counterions are located as a
point charge at $r=0$. Since they are spread over distances larger
than $\hx$, the behavior of the potential, although logarithmic, turns
out to have a different prefactor, see Eq. (\ref{eq:fin1}).

The case when $-1< Q/q<0$ is somewhat different. The effective interaction potential is no longer logarithmic and has a finite value at $x=0$. For $x\to0$,
\begin{equation}
  \beta V_{\eff}(x) = \beta V_{\eff}(0) + \frac{\hx^{2(1+Q/q)}}{
    \left(\frac{Q}{q}+1\right)\gamma(1+Q/q,N_b)} + O(\hx^2)
\end{equation}
with $V_{\eff}(0)$ given by Eq.~(\ref{eq:betaVeff0}). Notice that
since $0<1+Q/q<1$, the potential is again attractive at short
distances. Also the power law $\hx^{2(1+Q/q)}$ is
  different from the one of the case $Q/q>0$ where it was $\hx^2$. In
  this case, $-1< Q/q<0$, it is again possible to define the binding
  energy, that diverges when the limit
  $Q/q\to-1^{+}$ is approached. One can indeed show that $\beta
  V_{\eff}(0) \sim \ln \left(1+Q/q\right)$.

%%%%%%%%%%%%%%%%%%%%%%%%%%%%%%%%%%%%%%%%%%%%%%%%%%%%%%%%%%%%%%%%%%%%%%
\section{Conclusion and discussion}
\label{sec:concl}

We have introduced a classical system that exhibits some of the 
phenomenology at work in more complex colloidal suspensions. 
An ensemble of $N$ point particles with charge $q$ are free to
move within a disk of radius $R$, that bears a uniform background
charge of surface density $- q N_b/(\pi R^2)$. The corresponding
complex (mobile charges and background) forms a one component
plasma, with a global charge $(N-N_b)q$. A test point charge $Q$ is then
approached to the complex, perpendicularly to the disk plane,
along its axis of symmetry ($x$-axis, see Fig. \ref{fig:OCPdisk-charge}).
All charges were assumed to interact through a log potential, 
a choice that is convenient for the derivation of analytical results
and for the discussion of the physical mechanisms, but that we 
emphasized as somewhat irrealistic for a real Coulombic problem in three dimensions.
We have studied in detail the $x$-dependent effective potential $V_{\eff}$ experienced
by the intruder $Q$, defined as the free energy
of the complete charge distribution for a given distance $x$
between the test charge and the complex. 

At short distances $x$, $V_{\eff}$ is always attractive, with different
underlying mechanisms depending on the sign of $Q/q$. If the intruder 
and the mobile charges are like-charged, the intruder creates
its own correlation hole as it approaches the disk. The resulting short-range
attraction resulting from this polarization is analogous to its three dimensional counterpart
explaining charge inversion (overcharging, see Ref. \cite{GNS02}).
If on the other hand $Q/q<0$, the test particle attracts an excess of mobile
charges in the vicinity of the disk center, which overcomes the background - 
test charge repulsion. In this case, we found a diverging attraction 
for $Q/q\leq -1$, which precludes the definition of a binding energy
(cost to drag the test charge away from the disk, starting from 
$x=0$, the point of contact).

The long distance behaviour is also of interest.  If the complex has a
net charge, the leading contribution to $V_{\eff}$ reads $-Q q (N-N_b)
\ln \hx$, which leads to the expected like-charge repulsion at large
$x$. The neutral case $N=N_b$ is more subtle, and it has been shown
that a key quantity to rationalize $V_{\eff}$ is the quadrupolar
moment $\mathbb{Q}_{2}$ of the total charge distribution on the
disk. At large $x$, polarization effects disappear, and the mobile
charges adopt a profile that compensates for the background charge in
the bulk of the disk, while they are expelled from the immediate
vicinity of the disk edge $r=R$, thereby creating a charge imbalance
far from the disk center only. This necessarily leads to a negative
value of $\mathbb{Q}_{2}$, and hence to a repulsive behaviour at large
$x$, when $Q/q>0$. Indeed, what matters for large distance
interactions is the charges that are closest to the intruder, and they
happen to be the mobile charges expelled from $r=R$ (see the region
where $n>n_b$ in Fig. \ref{fig:ndens.xstar.infty}, for $N=50$ or
$N=500$, or equivalently, see the arrow in
Fig. \ref{fig:ndens-unbound} below).  The ensuing interaction is
repulsive when $Q$ and $q$ are of the same sign. This leads us to a
final remark that illustrates the subtlety of the long distance
effective potential. Consider a variant of the previous model, where
the mobile charges are no longer confined in the disk $r<R$, but can
explore the full disk plane (they are thus still 2D confined, but
unbounded in the plane). The uniform background, as before, is a disk
of radius $R$. We can repeat the analysis for $\Gamma=2$, which leads
to a profile $n(r)$ that departs from the one reported above in an
essential way: As can be seen in Fig. \ref{fig:ndens-unbound}, it is
monotonously decreasing, as happens to be the case at mean-field level
\cite{CMTR09} (i.e. for $\Gamma \to 0$).  For $N\geq N_b$, the decay
of the density profile $n(r)$ at large distances is algebraic in
$1/r^4$~\cite{J86,J03}, leading to a divergent
quadrupole. Furthermore, the density profile of this ``unbounded''
model shows a peculiarity, when $N\geq N_b$, $\int_{\mathbb{R}^2}
n(r)\,d^2\r=N_b-1$. Since there were originally $N$ mobile particles,
this means that $N-N_b+1$ particles have escaped to infinity. This can
be checked explicitly at $\Gamma=2$~\cite{J86,J03,FJT03}, but more
generally, it is a manifestation of the Onsager-Manning-Oosawa
condensation phenomenon~\cite{M69a, M69b, O71}: only a fraction
$(N_b-1)/N_b$ of the mobile ions are ``condensed'' inside or in the
vicinity of the disk. This is a consequence of the logarithmic
interaction between the ions and the disk when they are outside the
disk.  As a consequence, and at variance with the bounded model where
the charges stay in the disk, the global charge of the complex (disk +
mobile ions) is $-q$, whenever $N\geq N_b$. Therefore one expect that
the effective interaction of this complex with the charge $Q$ at large
$x$ will be attractive for $Q/q>0$. This situation is opposite to
the one met with the bounded model.

%% Increasing $\Gamma$, we speculate
%% that the following interesting scenario takes place: The mobile
%% charges will localize closer and closer to their equilibrium ground
%% state position, where all of them lie inside the disk (this can be
%% seen as a consequence of Earnshaw theorem \cite{E1842}, that states
%% that there is no stable equilibrium configuration with charges in
%% vacuum; hence the ground state requires that all $q$ ions are
%% surrounded by a portion of background). This means that for large
%% enough $\Gamma$, the differences between the ``bounded'' and
%% ``unbounded'' models become immaterial: the fact that the particles
%% are able or not to explore the region $r>R$ becomes irrelevant, as
%% confirmed by a recent 3D study of the plum pudding model
%% \cite{CCRT11}.  We are therefore back to the large $\Gamma$ results
%% reported in section \ref{sec:long-distance}, with a large distance
%% repulsion for $Qq>0$. Hence, in the unbounded model, the sign of the
%% effective force at a given large distance from the disk depends on the
%% temperature (or equivalently on $\Gamma$), while it is always of the
%% same sign in the ``bounded'' model.

\begin{figure}[htb]
  \null\vskip 6mm
  \centering
  \centering
  \includegraphics[width=\GraphicsWidth]{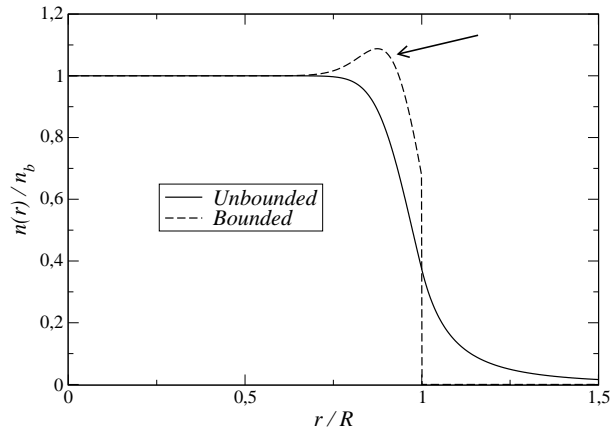}
  \vspace{5mm}
  \caption{\label{fig:ndens-unbound} Density profile of counterions in
    the disk, for the ``unbounded'' and ``bounded'' models
    (counter-ions are either allowed to explore the region $r>R$, or
    not).  Here $N=N_b=40$ and $\Gamma=2$. The counter-ion excess for
    the bounded model --shown by the arrow, and studied extensively in
    this paper-- leads to a negative quadrupole moment
    $\mathbb{Q}_{2}$, see section \ref{sec:long-distance}, while in
    the unbounded case, one mobile ion escapes to infinity, leaving
    the complex (disk + ions) with a net charge $-q$. This results in
    large distance effective forces on the test charge that have
    opposite signs.}
  \vspace{6mm}
\end{figure}

%%%%%%%%%%%%%%%%%%%%%%%%%%%%%%%%%%%%%%%%%%%%%%%%%%%%%%%%%%%%%%%%%%%%%%

We would like to thank L. \v{S}amaj for interesting discussions,
and for having provided us with some of the coefficients $c_\mu$ required
to compute the partition functions in section \ref{ssec:arb}.  The
support of ECOS-Nord/COLCIENCIAS-MEN-ICETEX is also gratefully
acknowledged. G.~T.~acknowledges partial finantial support from Comit\'e de Investigaciones y Posgrados, Facultad de Ciencias, Universidad de los Andes. 

\bibliographystyle{apsrev}
\bibliography{biblio}

\end{document}